\documentclass[11pt]{article}

\usepackage[a4paper,margin=1in]{geometry}
\usepackage[T1]{fontenc}
\usepackage[utf8]{inputenc}
\usepackage{lmodern}
\usepackage{graphicx}
\usepackage[hidelinks]{hyperref}

\title{Social Simulations: from Agent-Based Modeling to Digital Twins}
\author{
Erica Cau\\
\small Department of Computer Science, University of Pisa\\
\small \texttt{erica.cau@phd.unipi.it}
\and
Andrea Failla\\
\small  ISTI-CNR\\
\small \texttt{andrea.failla@isti.cnr.it}
\and
Valentina Pansanella\\
\small ISTI-CNR\\
\small \texttt{valentina.pansanella@isti.cnr.it}
\and
Giulio Rossetti\\
\small ISTI-CNR\\
\small \texttt{giulio.rossetti@isti.cnr.it}
}
\date{}

\begin{document}

\maketitle

\begin{abstract}
Social simulation is a computational approach for studying how social phenomena emerge from interactions among individuals, groups, organizations, and their environments. This article surveys the progression from classical agent-based modeling to large-language-model-enhanced agent-based simulations and social digital twins. It outlines their core modeling assumptions, architectural components, advantages, limitations, and the scope of claims that these approaches can support.
\end{abstract}

\section*{Synonyms}
Agent-based Modeling, Generative Agent-Based Modeling, LLM-based Agent-based Modeling, Digital Twin, Social Digital Twin.

\section*{Glossary}
\begin{itemize}
    \item \textbf{Agent-Based Modeling}: A computational method that simulates the interactions of autonomous agents to study the emergent behavior of complex systems.
    
    \item \textbf{Generative Agent-Based Modeling/LLM-based Agent-Based Modeling}: A form of agent-based modeling where large language models drive agent reasoning, communication, and decision-making in simulations.

    \item \textbf{Digital Twin}: A virtual representation of a physical system, process, or entity that mirrors its real-world counterpart.

    \item \textbf{Social Digital Twin}: A digital twin designed to model and simulate social systems, human interactions, and collective behavior to analyze and predict societal dynamics.


    \item \textbf{Socio-Technical Systems}: Systems composed of interacting social and technical elements, where human behavior, organizational processes, and technological infrastructures are deeply interdependent and jointly shape overall system outcomes.

\end{itemize}

\section{Definition}
\label{sec:definition}
Social simulation is the computational study of social phenomena through the construction and analysis of formal models that represent individuals, groups, organizations, and their interactions, with the aim of generating artificial societies \textit{in silico} to observe, explain, or predict emergent patterns of collective behavior. It is the algorithmic continuation of classical thought experiments, translating stylized “what-if” scenarios of social theory into executable models whose outcomes can be observed, measured, and systematically varied.

\section{Social Simulations}
\label{sec:introduction}
Social systems are inherently complex: collective outcomes emerge from the interactions of heterogeneous individuals, whose behaviors, beliefs, and relationships evolve over time. Traditional analytical methods often struggle to capture this complexity, as they rely on simplified assumptions or static snapshots that cannot represent dynamic, adaptive social processes.
\textbf{Social simulation} is a computational approach that addresses this challenge by creating artificial societies \textit{in silico} in which agents interact according to explicit rules. These simulations allow researchers to observe how micro-level behaviors generate macro-level phenomena, to test theoretical assumptions, and to explore counterfactual “what-if” scenarios that are difficult or impossible to experiment with in the real world.
At its core, social simulation is the algorithmic continuation of classical thought experiments in social theory. Philosophers and social scientists have long employed stylized “what-if” reasoning to explore social phenomena: for example, how segregation might emerge even if individuals have mild location preferences, or how norms stabilize in small communities. Social simulation formalizes these thought experiments into executable models, producing measurable outcomes that can be systematically analyzed.
The field draws on diverse methods, including agent-based modeling (ABM), system dynamics, network simulations, and, more recently, AI-enhanced models and digital twins. ABMs, in particular, provide a natural bridge between theory and computation by representing individual agents with distinct attributes, states, and behavioral rules within structured environments. These models make it possible to study emergent dynamics such as cooperation, polarization, diffusion of innovations, and social influence in a controlled, repeatable manner.
In this chapter, we first introduce classical ABMs as the foundation of social simulations, highlighting their structure, design, and limitations. We then present advances in ABM enhanced with Large Language Models (LLMs), which enable agents to interact using natural language and simulate complex cognitive and social processes. Finally, we discuss the integration of social simulations into digital twins, allowing high-fidelity, data-driven representations of real-world socio-technical systems. Together, these perspectives illustrate the evolution of social simulation from conceptual experiments to computational frameworks capable of addressing pressing questions in social science.

\subsection{Classical ABM}
\label{subsec:abm}
Agent-based models (ABMs) 
are a form of \textit{computational social science} \cite{conte2012manifesto}, involving models implemented as computer programs \cite{gilbert2019agent} that represent social systems as collections of autonomous, heterogeneous agents interacting according to explicit rules within structured environments \cite{bianchi2015agent}. 
ABMs formalize theoretical assumptions and explore how micro-level interactions produce emergent collective dynamics, such as cooperation, segregation, polarization, or diffusion \cite{bianchi2015agent}. 
\\ \ \\
\textbf{\textit{The Architecture.}}
The \textbf{design} of an ABM involves designing computational experiments that can address one or more research questions through the agents' local behaviors. 
These questions determine the specific social mechanism 
under investigation, whether it concerns opinion polarization, disease transmission, market dynamics, or urban segregation.
The choice of a specific research question drives the definition of the model and the design of the initial population. 

\smallskip
\noindent \textbf{Agents} constitute the fundamental building blocks. 
An agent represents an abstraction of a real-world entity -- an individual, organization, or other -- that possesses agency, meaning the capacity to make decisions and take actions. 
Each agent is characterized by three primary elements: attributes, states, and rules. 
\begin{itemize}
    \item \textit{Attributes} are agent-level variables that represent relatively stable characteristics that define the identity of agents and, collectively, of the population. 
    These may include demographic variables such as age, gender, or education level, economic variables, such as the income level or job title, as well as psychological traits such as risk tolerance or cognitive biases.  
    Attributes usually remain constant throughout a simulation, or change at a much slower rate than other agent properties. 
    Their evolution is not the \textit{objective} of the study; rather, the modeler wants to see their causal impact on the dynamics. 
    In many classical models, attributes are assumed to be homogeneous across the population; however, one of the main strengths of ABMs lies in their ability to introduce heterogeneity among agents.
    Attributes are normally part of the \textit{model parameters}, i.e., the set of variables that influence the studied dynamics. 
    This is possible because they normally become part of the \textit{rules} governing agents' behavior. 
    In this way, modelers can evaluate their causal impact on the emergent behaviors that arise from the social simulations. 
   
    \item \textit{States} represent the current condition of an agent and are the properties that change throughout a simulation. 
    Their evolution is driven by the model's \textit{rules}, which update states over time based on interactions among agents and with the environment. 
    Because states capture what changes, they are often the focus of analyses of emergent phenomena. 
    For example, an agent's \textit{opinion} can be modeled as a state, allowing researchers to study the emergence of polarization or consensus; similarly, in epidemic models, states such as \textit{susceptible}, or \textit{infected}, enable the study of how diseases spread through a population. 
    States can be binary, discrete, or continuous variables, as well as more complex mathematical entities. 
    
    \item \textit{Rules} define how agents behave within the model. In practice, this means that each agent follows an internal algorithm that guides its choices, determines its actions and interactions, and updates its \textit{states}. 
    Rules are usually grounded in cognitive or sociological theories, 
    and must be explicitly specified and logically consistent, even when they incorporate stochastic elements. 
    Through rules, specific behavioral mechanisms are connected to observable outcomes, enabling researchers to test theoretical propositions about how individual behaviors generate collective patterns. When a rule involves an agent's attribute, it becomes an \textbf{agent-level parameter}.
    
\end{itemize}

\smallskip \noindent Rules can also be influenced by \textbf{model parameters}, which are properties defined at the population level. 
These parameters affect the behavior of all agents in the same way and are conceptually distinct from agent-level attributes, which can vary across individuals. 
Examples include the bias of a recommender system that curates agents' timelines or the transmission probability of a virus in an epidemic model. 

\smallskip \noindent In social systems, much of the observed complexity arises from interactions among agents, which are usually constrained by the relationships that connect individuals. 
For instance, it is more likely that one's opinion will be influenced by friends, colleagues, or family members than by complete strangers. 
\textbf{Relationships} between agents are commonly represented using network structures, most often graphs, where agents are modeled as nodes and their relationships as edges. 
Classical network models (e.g., random, small-world, or scale-free graphs) often provide a framework for the design of agents' societies. 
The topology of such a network is a key determinant of system-level outcomes, influencing diffusion, coordination, and emergent behaviors.
Like agents, relationships can also have \textit{attributes}.
These relationships can be static or dynamic. 
As state dynamics, relationship dynamics are also governed by rules. When relationships evolve over time, dynamic networks provide a natural formalism for representing them.
More advanced ABMs use multi-layer networks, hypergraphs, or more complex structures that allow different types of relationships (e.g., family, friendship, professional ties) to be represented simultaneously. 
Finally, relationships can also be adaptive. 
In many domains, the evolution of social ties and the evolution of agent states are interdependent. Adaptive networks, also called co-evolving networks, provide a modeling framework in which both the network structure and the states of its nodes change over time. 
This co-evolution can generate complex behaviors and emergent patterns that would not arise if either the network or the agent states evolved independently. 

\smallskip \noindent \textbf{Time} represents a critical but challenging aspect of Agent-based modeling. 
Real-world social processes occur in continuous time, but computational models must discretize time into manageable steps. 
Modelers must decide on the appropriate time step duration, which determines how many actions or interactions occur within each simulation cycle. 
Additionally, they must choose between synchronous updating (where all agents act simultaneously) and asynchronous updating (where agents act sequentially), as this choice can significantly impact simulation outcomes due to path-dependence effects.

\smallskip \noindent \textbf{Space} provides another dimension of complexity in agent-based models. 
While some models operate in abstract (or virtual) social spaces defined only by relationships, others incorporate explicit spatial representations. 
Spatially explicit models range from simple grid-based environments to complex geographic information system integrations that incorporate real-world spatial data. 
The spatial component becomes particularly important when modeling phenomena such as urban development or epidemic spread, where physical proximity and environmental characteristics play crucial roles.

\smallskip \noindent
Besides defining the model and the initial population, where agents, relationships, time, and space are specified, the \textbf{design} of an agent-based social simulation also involves several additional decisions like choosing the initial state distribution for agents, which may be uniform, biased, or based on empirical data; selecting the range of parameter values to explore in order to test their effects on the system; determining the duration of a single simulation run, either in terms of the number of actions or interactions, discrete time steps, or until a stopping condition is met; and finally, identifying which metrics to compute during each simulation run, so that results can be aggregated across runs to produce meaningful experimental outcomes.
\\ \ \\
\textbf{\textit{Agent-based Social Simulations.}}
Agent-based modeling in the social sciences emerged in the 1960s–70s, with the seminal work on segregation by Thomas Schelling \cite{schelling1971dynamic} introduced simulations of individual behavior and interaction, as an alternative to dominant macro-level system theories.

The 1990s marked the peak of ABM popularity, thanks to increasing computational power and the development of dedicated computer platforms, e.g., NetLogo \cite{wilensky1999netlogo}. 
An advantage of these platforms is that they allow even non-computer scientists to develop models, enabling the development of a multidisciplinary field of research at the intersection of sociology, psychology, computer science, mathematics, physics, and complex systems theory \cite{bianchi2015agent}. 
Two of the most prominent examples from this decade of work are Axelrod’s dissemination of culture \cite{axelrod1997dissemination}, and Epstein and Axtell's Sugarscape \cite{epstein1996growing}. 
Now, several programming languages allow for the creation of agent-based models, also through specialized libraries \cite{rossetti2018ndlib, terhoeven2025mesa}. 

More recently, Agent-based social simulations have been applied in economics and finance \cite{axtell2025agent}, sociology \cite{bianchi2015agent}, and other fields.

\noindent Agent-based models where states are \textbf{binary} are particularly well-suited for studying processes where the phenomenon of interest can be characterized as something that agents either \textit{possess} or \textit{do not possess}. 
These models typically employ a contagion metaphor where agents transmit states to one another through social influence mechanisms. 
In the modeling of the diffusion of innovations \cite{kiesling2012agent}, agents transition from non-adopters to adopters of new technologies or practices. 
Information diffusion \cite{zhang2019empirically} models study how news, rumors, or knowledge spread through social networks. 
In recent years, researchers have focused specifically on online social networks \cite{guille2013information}. 
Following this approach, researchers have approached diffusion of behaviors \cite{prabhakaran2023improving} and emotions \cite{van2023emotion}. 
Voting behavior models \cite{holley1975ergodic} represent another important application area, where agents' electoral choices are influenced by their social interactions.
Epidemiological models \cite{lorig2021agent}, while not exclusively social in nature, represent a significant application of discrete state ABM in studying disease spread through social networks.
Agent-based models where states are \textbf{continuous} address phenomena that exist on spectrums rather than as binary states. Models of opinion dynamics \cite{sirbu2016opinion} represent the most prominent application in this category, studying how individual opinions evolve through social influence. 
\\ \ \\
\textbf{\textit{Advantages and Limitations.}}
Agent-based models offer several distinctive \textbf{advantages} for studying social phenomena. 
They naturally capture complexity and nonlinearity, accommodating heterogeneous agents with diverse behavioral rules and interaction patterns. 
ABMs are particularly suited to studying emergent phenomena, showing how simple micro-level rules can generate macro-level patterns such as convergence, polarization, or segregation. 
Their transparency is another strength: behavioral assumptions are explicitly coded, making mechanisms open to inspection and testing, unlike many statistical or machine learning models. 
A distinctive advantage is the ability to study \textit{causal relationships}. 
While empirical data often reveal only correlations, ABMs embed mechanisms directly, allowing researchers to vary parameters systematically and identify their causal effects on target outcomes. 
Finally, ABMs serve as effective platforms for policy analysis, enabling scenario testing of interventions aimed at individual behaviors and their population-level consequences, such as in health or environmental policies.  
\smallskip \noindent

ABMs also face important \textbf{limitations}. 
They are built on assumptions about behavioral mechanisms that, though theory-informed, may be arbitrary, without validation. 
In opinion dynamics, for example, models often assume opinions converge through averaging, yet alternative mechanisms (e.g., cumulative effects, backfire) are equally plausible. 
Network analysis and network mining techniques play a crucial role in informing, calibrating, and validating agent-based models.
Empirical network data can be mined to infer realistic interaction structures, estimate model parameters, and identify relevant mesoscopic patterns such as communities or motifs.
Conversely, networks generated by simulations can be analyzed using standard network metrics to compare simulated and real-world systems, thus closing the loop between modeling, simulation, and empirical observation.
However, there is a lack of a robust validation framework: outputs may resemble real data without ensuring that the underlying mechanisms are correct, and most models are not calibrated with empirical initial conditions. 
As a result, many ABMs function better as thought experiments than as tools for direct application, whether for prediction, inference, or policy-making. 
As models grow more complex, their analysis can become as difficult as the study of empirical data itself, limiting their explanatory power. 
Finally, computational demands can be significant, especially in large or detailed models, raising both technical and environmental concerns. 

\subsection{ABM Enhanced with AI}
\label{subsec:2}
The rise of artificial intelligence, especially \textbf{generative agents}, has introduced a new class of computational models that bridges classic ABMs and social digital twins. 
This evolution marks a shift in social simulations, moving from traditional rule-based ABM toward \textit{LLM-enhanced agent-based modeling}, in which agents produce context-aware arguments that mirror human discourse patterns.
This approach addresses one of the main limitations of classical agent-based models: while the traditional model relies on mechanistic rules for modeling the actions of agents that do not consider the linguistic dimension necessary in social interaction, LLM-based agents can produce emergent behaviors that arise, in a natural way, from conversational dynamics and contextual interpretation. 
In this way, not only is it possible to understand the interaction patterns of the \textit{nodes} of the network, but also to understand the way arguments and conversations unfold to capture complex social behaviors, such as persuasion and argumentation strategies of LLMs, and the impact of psychological biases. 
For these reasons, LLM-based agent simulations might impact other fields beyond computational social science itself. 
Such simulations can help understand how these models express stances, construct arguments, and, more broadly, whether they are capable of inferring others' internal states and adapting their responses, approximating a Theory of Mind as an emergent cognitive capability of LLMs.
\\ \ \\
\textbf{\textit{Architecture of LLM-ABM.}}
Although still in its early stages, this research spans a wide variety of approaches. 
An LLM-enhanced agent-based simulation primarily consists of a population of \textbf{agents} and an underlying \textbf{language model}, such as Llama or Mistral, that generates each agent's reasoning and comments. 
Agents are prompted with a specific role or behavior (e.g., one agent is tasked with persuading the other). 
At the same time, the \textbf{debate topic} and the number of \textit{interactions} define the simulation's context and settings.
Distinct personalities or divergent initial opinions may further characterize agents through prompts. 
They can also be equipped with memory mechanisms, allowing them to retain past interactions and adapt future responses accordingly.
Starting from this minimal configuration, additional components can be introduced by drawing on classical ABMs. 

Agents can interact in a mean-field scenario or be embedded in a network structure tailored to the study's aim (e.g., small-world or scale-free). 

Other extensions include integrating interaction rules or opinion update rules from the ABMs literature to regulate when interactions occur and how opinions evolve. In this way, agents may be influenced by their neighbors, adjust their stance based on confidence bounds, or follow inertia or peer pressure. 
This approach enables a more thorough investigation of the interplay between social ties and natural language arguments, potentially yielding new insights into how information spreads within networked populations.
\\ \ \\
\textbf{\textit{Studies on LLM-opinion dynamics.}}
Current applications of LLM agent simulations have explored a range of social phenomena traditionally studied in \textit{computational social science}, including \textit{opinion dynamics}, \textit{persuasion} mechanisms, and \textit{polarization}. 
Early studies show that LLM agents can demonstrate structured behavior, such as organizational skills, relationship-forming capabilities, and coordination of group activities \cite{park2023generative, park2022social}.

In the field of opinion dynamics, LLM agents have been employed to simulate debates in which individuals exchange arguments in natural language. 
In mean-field interaction settings, they tend toward agreement through structured persuasion, in which agents reach consensus by selectively adopting peers’ arguments rather than blindly accepting opinions \cite{cau2025selective}.
Instead, when embedded in \textbf{network structures}, LLM agents can interact and/or change their opinions according to rules derived from classical opinion-dynamics models.  
Applying bounded-confidence interaction rules promotes polarization and the formation of echo chambers. 
Simulation results show that small-world and scale-free networks encourage echo chamber formation and the formation of large clusters of like-minded agents, mirroring real-world patterns of information diffusion \cite{wang2025decoding}.
Within this same networked setting, polarization can be further induced by prompting agents with \textbf{cognitive biases}.
For example, simulated confirmation bias affects how agents interpret incoming arguments, with stronger biases influencing agents to maintain polarized positions and weaker biases increasing opinion adaptability during interaction \cite{chuang2024simulating}.
\\ \ \\
\textbf{\textit{Advantages and limitations.}}
LLM-based simulations remain structurally similar to classical ABMs: the \textbf{agents} interact locally through \textbf{network ties} and generate \textbf{emergent} collective patterns at the \textbf{population} level. Their main advantage lies in their language-mediated interactions, which allow LLM-agents to engage in argumentation and persuasion and to make micro-level influence mechanisms more evident through language. 
Moreover, LLM-based ABMs allow us to investigate \textit{in vitro} scenarios to determine whether LLM-agent populations spontaneously exhibit characteristics similar to those of human populations, and to explore scenarios that would be infeasible to test in real-world settings. In this sense, LLM-based simulations can support \textbf{explanatory} claims about how specific interaction mechanisms generate collective outcomes, and \textbf{exploratory} investigations of plausible social scenarios under hypothetical conditions.

At the same time, LLM-based simulations face important limitations. LLMs are statistical models trained on incomplete datasets, which might reflect cultural and demographic biases. Simulation outcomes are highly sensitive to prompt design, raising concerns about robustness and applicability, especially given the stochastic nature of their outputs. 
Many current studies also rely on simplified or absent network structures, overlooking a key driver of real-world social influence.
Validation presents additional issues. 
Unlike traditional opinion dynamics models, the presence of linguistic interactions makes benchmarking less straightforward and increases the risk of mistaking model biases for realistic social phenomena. A common approach is to compare the LLM-enhanced simulation with corresponding agent-based models to understand the patterns introduced by the LLM.
As a consequence, LLM agent models cannot provide reliable \textbf{predictive} claims about real-world systems, and their realistic language should not be considered as evidence of the ability to properly simulate human behavior.
A promising approach is the development of \textit{social digital twins}, which can ground LLM-based simulations in richer and more realistic representations of social systems.

\subsection{Digital Twins}
\label{subsec:3}
So far, we have touched on traditional and AI-enhanced Agent-based social simulations.
What these two paradigms have in common is that they are \textbf{task-oriented}: the primitives of the simulation are designed specifically to answer a defined research question, or to solve a particular task.
In a generic epidemic model, for instance, the objective is to understand how a pathogen spreads and (possibly) dies out.
As a result, agents are equipped only with the minimal set of attributes necessary for that task --- typically, their current state and transition probabilities.
\\
Digital twins adopt a different stance, bringing simulations one step closer to realism.
A digital twin (DT) is a virtual replica of a specific physical, biological, or social system, designed to mirror its structure, state, and behavior using empirical data~\cite{barricelli2019survey}. 
Instead of tailoring the model to a narrow problem, DTs aim to replicate an entire, living system \textit{in silico}, preserving its complexity and detail, even when such detail is not immediately relevant to a single research objective.
The goal is thus not simply to approximate behavior under specific assumptions, but to construct a dynamic, data-driven representation of a particular system capable of supporting multiple forms of investigation.
\\ \ \\
As such, each DT is anchored to a \textbf{real-world referent}, that is, the object or system it is replicating. 
For instance, one does not build the DT of a generic city, but of a specific one — e.g., the DT \textit{of the city of Rome}. 
This twin is directly tied to its real-world counterpart and may ingest live data from it, such as traffic, mobility, infrastructure, and human behavior data.
Depending on the nature of the system they aim to replicate, DTs can be ascribed to one of three main categories:
\begin{itemize}
    \item DTs of \textit{mechanical systems}, which replicate tangible, material systems governed by physical laws (e.g., the DT of a car, the DT of a bridge).
    These are typically coupled with the source system through real-time sensors that continuously capture and feed this data into the digital replica.
    \item DTs of \textit{biological and cognitive systems}, which model individual living organisms and their physiological and/or psychological functions. 
    These twins aim to reproduce the internal states of biological entities like humans (e.g., DTs of medical patients, DTs of athletes);
    \item DTs of \textit{socio-technical systems}, which represent interacting agents embedded within technological and/or social infrastructures. 
    While the previous category focused on mirroring individual living entities, under this category fall all those DTs that capture the collective dynamics emerging from the interactions among multiple agents and their environment (e.g., DTs of hospitals, DTs of social media platforms).
\end{itemize}

\noindent Clearly, the third category is the most relevant to the domain of social simulations. 
Socio-technical digital twins (SDTs) usually retain the core ABM paradigm but reduce its abstraction and also further ground the paradigm in real data.
This epistemic switch ultimately redefines the question behind social simulation studies: from ``what happens, in general, to societies with these characteristics?'' to ``what will happen here, now, under these conditions?''.
\\ \ \\
In practice, SDTs offer multiple advantages with respect to traditional simulative approaches.
First, they provide a \textbf{controlled environment} for testing counterfactual scenarios that would be costly or unethical in the real world. 
For example, a digital twin of a social-media platform allows researchers to vary recommendation algorithms and observe their effects on network-mediated processes without exposing actual users to untested interventions. 
Re-running the twin under different settings yields comparable trajectories for systematic evaluation~\cite{rossetti2024social}.
Similarly, a DT of urban mobility can simulate how changes in transportation policies reshape commuting networks, while a DT of an organization can explore how modifications in communication structures affect productivity or knowledge diffusion.
\\
Second, SDTs are \textbf{reusable}: once the core model is calibrated, the same infrastructure can support multiple investigations; the same SDT of a social media platform allows studying recommendation impact, misinformation diffusion, echo-chamber formation, or other processes, all without rebuilding the model.\\
High-fidelity grounding also yields \textbf{augmented forecast granularity}.
While traditional ABMs typically return aggregate trends (e.g., the expected size of an information cascade) a SDT can generate micro- and meso-level predictions. 
\\
Finally, realistic SDTs can generate \textbf{substitute data} when direct access is limited. 
The recent tightening of social-media APIs, for instance, restricts the availability of interaction logs, but a digital twin can produce synthetic datasets that approximate the missing evidence. 
\\ \ \\
Yet, as with any technological promise, the adoption of SDTs comes with challenges.
First and foremost are the \textbf{costs}: developing and maintaining a high-fidelity SDT is resource-intensive.
It requires fine-grained data, computational infrastructure, and deep domain expertise.
But the more fundamental issue is \textbf{validation}.
To be reliable tools for reasoning and prediction, SDTs must be validated on at least three levels.
First, on a \textit{structural} level, researchers should assess whether the architecture of the digital twin accurately reflects the composition and organization of the real-world system it replicates.
This includes validating the network structure, agent types, affordances, and constraints, ensuring that the components and their relationships correspond to those observed in reality.
Second, on a \textit{behavioral} level, it is essential to verify that the agents within the twin behave in plausible and heterogeneous ways.
This means not only reproducing average trends but capturing the diversity of strategies and responses that emerge in the real system.
In social systems, especially, overlooking behavioral heterogeneity can indeed lead to misleading forecasts.
Third, on a \textit{predictive} level, the outputs of the SDT should be compared with empirical observations to evaluate its forecasting power.
This does not imply perfect correspondence at all times, but the twin should be able to reproduce past or ongoing dynamics with reasonable accuracy under known conditions.
\\ \ \\
Finally, greater realism is not inherently an epistemic virtue. While high-fidelity models like SDTs allow more granular insights and predictions, they also risk overfitting to the peculiarities of the system they emulate.
In this sense, realism can become a double-edged sword, and researchers should carefully weigh the trade-offs between specificity and abstraction. 
The choice between a digital twin and a more abstract ABM should depend on the research goals: whether the priority is to forecast outcomes in a particular system, or to uncover general mechanisms that apply across contexts. 

\noindent More generally, the scope of the claims supported by SDTs and LLM-based simulations must be interpreted with care.
These models can reliably support \textbf{exploratory} uses (generating hypotheses, testing counterfactuals) and \textbf{explanatory} uses, by identifying mechanisms that could account for observed dynamics.
Their \textbf{predictive} use, however, is necessarily more limited: forecasts are conditional on data quality, calibration, and behavioral assumptions, and should be regarded as scenario-dependent projections rather than precise point predictions.
Over-interpretation is a concrete risk, especially when high-fidelity outputs create an illusion of accuracy; apparent realism does not guarantee epistemic validity.
Consequently, results from SDTs and LLM-driven agents are best understood as decision-support tools that inform reasoning about possible futures, not as definitive or context-independent predictors.

\bibliographystyle{plain}
\bibliography{bibliography}

\end{document}